\documentclass[letterpaper]{mn2e}

\def\simpropto{\lower.2ex\hbox{$\; \buildrel \propto \over \sim \;$}}
\def\ltsim{\lower.5ex\hbox{$\; \buildrel < \over \sim \;$}}
\def\gtsim{\lower.5ex\hbox{$\; \buildrel > \over \sim \;$}}

\usepackage{multirow}
\usepackage{rotating}
\usepackage{colortbl}
\usepackage{color}
\usepackage[fleqn]{amsmath}
\usepackage{amssymb}
\usepackage{amsfonts}
\usepackage{verbatim}
\usepackage{scalefnt}
\usepackage[percent]{overpic}

\def\gtorder{\mathrel{\raise.3ex\hbox{$>$}\mkern-14mu
     \lower0.6ex\hbox{$\sim$}}}
\def\ltorder{\mathrel{\raise.3ex\hbox{$<$}\mkern-14mu
     \lower0.6ex\hbox{$\sim$}}}

\newcommand{\beq}{\begin{equation}}
\newcommand{\eeq}{\end{equation}}
\newcommand{\ba}{\begin{eqnarray}}
\newcommand{\ea}{\end{eqnarray}}
\def\spose#1{\hbox to 0pt{#1\hss}}

\newcommand{\comments}[1]{} 

\title{From diffuse extragalactic and galactic gamma rays to limits on extra dimensions} 

\author[Cass\'e]{Michel Cass\'e$^1$, Bruno  Mansouli\'e$^1$, Joseph Silk$^{2,3,4}$\\
$^1$ IRFU, CEA, University Paris-Saclay, Gif-sur-Yvette, France\\
$^2$ Institut d'Astrophysique de Paris, CNRS, 98bis bd Arago, 75014 Paris, France\\
$^3$ Dept of Physics  \& Astronomy, The Johns Hopkins University, Baltimore MD 21218, USA\\
$^4$Beecroft Institute of Particle Astrophysics and Cosmology, 
Department of Physics, University of Oxford, 1 Keble Road Oxford, OX1 3RH, UK\\
}
\begin{document}

\maketitle

\begin{abstract}

We derive the maximum fraction of energy emitted in the form of massive (Kaluza- Klein) gravitons by core collapse supernovae, and the corresponding minimal extra-dimensional Planck mass M* in the ADD gravity framework at TeV-scales. Our constraints arise: a) from the extragalactic gamma ray background observed by Fermi- LAT after astrophysical sources have been removed, and b) via the residual galactic emission left after astrophysical and potentially dark matter emission have been removed. We focus on a number of extra dimensions 3 and 4, since M* is then in the TeV range, where astrophysical and collider constraints compete.  Lower limits on M* are derived in  case a) of 8.0 TeV and 1.1 TeV, and in case b) of 16 TeV and 1.9 TeV,  for a number of extra dimensions n=3 and n=4 respectively.  These limits are especially robust and insensitive to the various uncertainties involved. 

 \end{abstract}

 \section{Introduction\label{sec-intro}}
The standard model of particle physics is  compelling but it fails to explain why gravity is so much weaker than the electromagnetic or nuclear forces. The huge difference in strength of fundamental forces is one aspect of the "hierarchy problem"; another  is the Higgs boson mass which is much lower than it should be in  the absence of supersymmetry.
This is manifested by the difference between the electroweak and gravitational energy scales (about 1TeV against $10^{16} $ TeV) and  the difficulty of its stabilisation  \cite{2012EPJC...72.2020B}. 
 Hidden extra dimensions have been invoked to dilute gravity and make it appear feeble  \cite{1999PhRvD..59h6004A,1999PhRvL..83.3370R}
 whereas in the extradimensional realm, gravity  is as strong as electromagnetism. 

In the following, we adopt the ADD framework  \cite{1999PhRvD..59h6004A} expected to give a purely geometric explanation to the hierarchy puzzles (why is G so feeble? why is the Planck mass so high?). In this framework, inspired by the Kaluza-Klein hypothesis of the existence of compact extra dimensions (ED), quantum gravity features providentially  at about the TeV energy scale, which renders it accessible to experimental
and observational verification through the transient existence  of massive, unstable, Kaluza-Klein gravitons (hereafter referred to as gravitons). This bold hypothesis takes the risk of  being tested  physically in the laboratory (table-top experiments, via spectroscopy  at CERN, and astrophysically through gamma ray astronomy. Cosmological constraints are stringent \cite{2001PhRvL..87e1301H} 
but more uncertain than those discussed here.
We derive robust limits on the fundamental  (extra-dimensional) Planck mass M* for a number of ED, $n= 3 $ and 4, from gamma ray observations, and compare these results  with those derived from LHC experiments. For M*, we adopt the convention of \cite{2003PhRvD..67l5008H}, defining M* through the formula $M_{Pl}/(8\pi) = (2\pi R)^n (M^*)^{n+2}$.

Note that ED do not  "explain" the weakness of gravitation but geometrize the puzzle. 
Thus, if we assume that  large ED exist in nature, they would inevitably have been present in the past, but today are compactified. However at high energies, there is a potential signature via the generation of gravitons which would be emitted in supernova (SN) cores with energies around 100 MeV, providing a significant contribution to the celestial diffuse gamma ray background via radiative decays. This potential contribution allows us to set independent limits from those deduced from individual neutron stars, around which gravitons are trapped and decay into gamma rays \cite{2001PhRvL..87e1301H,2002PhRvL..88g1301H},  
most recently constrained by Fermi data \cite{2012JCAP...02..012F}
as well as by  neutron stars grouped in the galactic bulge  \cite{2004PhRvL..92k1102C}.
High energy neutrino bounds have also been examined but are generally weaker  \cite{2018PhRvD..98l3009A}.
 In this Letter, we derive the maximum fraction of energy emitted in the form of gravitons by all core collapse SN, failed or not, and  assess the corresponding minimal extra-dimensional Planck mass. Our limits are constrained by both the extragalactic isotropic gamma ray background and the diffuse galactic emission, less galactic centre excess associated either with  dark matter annihilations or millisecond pulsars    observed by the Fermi LAT after astrophysical sources have been removed. We focus on a number of ED 3 and 4, since the fundamental Planck mass is then in the TeV range, where astrophysical and collider constraints compete. Its value is robust and insensitive to the various uncertainties involved. 

\section {EXTRA DIMENSIONS AND THE ISOTROPIC GAMMA RAY BACKGROUND }
We focus on the isotropic extragalactic gamma ray background produced by all SN from z = 0 to $z \sim  5, $ updating the seminal earlier  work \cite{2001PhRvL..87e1301H} 
and
assuming, like the authors, a toroidal compactification of ED, all with the same radius. In general,   during core collapse, matter reaches such a high average temperature ($\sim$ 30 MeV) that gravitons could be abundantly produced. Due to the extreme weakness of the gravitational force, they escape the dense stellar core, unimpeded. Each graviton of mass m subsequently decays in space into two gamma rays, of m/2 energy, producing a diffuse gamma ray background that could be observable for large enough ED or sufficiently small extra-dimensional Planck mass M*. Since this is not the case, we derive a lower bound on M*.

The value of M* is proportional to $f_{KK}^{-1/(n+2)}$  \cite{2002PhRvL..88g1301H}, where $f_{KK}$ is the fraction of energy  emitted in the form of Kaluza-Klein gravitons during the core-collapse of a SN. 
Hence its value is especially robust and insensitive to the various uncertainties involved in this method, most notably mean temperature during core collapse, equation of state of superdense matter, stellar evolutionary model, explosion simulation, SN rate, model of galactic evolution, and fraction of the gamma ray background left in photons from graviton decays. 

We take advantage of detailed calculations of the extragalactic antineutrino flux arising from SN (successful or failed) based on hydrodynamical simulations of the Garching group \cite{2018JCAP...05..066M,2017JCAP...11..031P} 
(and references therein) along with  reasonable models of galactic evolution to derive the hypothetical Isotropic Gamma Ray Background (GRB) arising from graviton decay and compare it to the residuals left by astrophysical sources. Indeed, the estimate of \cite{2001PhRvL..87e1301H} 
based on the GRB produced by all SN since the creation of the first stars is excessively conservative (no evolution of the SN rate as a function of redshift, no failed SN, all the observed emission not surpassed by graviton decay). It is better to use the residuals left after all the contributions of astrophysical sources have been removed, which is  good practice in  searches for dark matter signatures. Indeed, little room is left for exotic processes (dark matter, primordial black holes, quasars and blazars) since the GRB is well explained by the gamma ray emission from galaxies, quasars and blazars  \cite{2016PhRvL.116o1105A}

The energy of a typical SN is $3\times10^{53}$ ergs or $1.8 \times  10^{59}$  MeV and the number of electron antineutrinos released (with a mean energy of $\sim$15 MeV) is  $2\times10^{57}$. If all the energy is emitted in the form of gravitons ($f_{KK}$ = 1) with an energy of about 100 MeV (see Table 2 in \cite{2003PhRvD..67l5008H}), the number of gravitons emitted is 15\% that of antineutrinos, and the corresponding gamma ray emission is 30\%, since each graviton decays into two gammas. Empirically, there is little room left to non-astrophysical sources (dark matter annihilation or decay and/or graviton decay.) The maximum gamma ray flux left for exotic sources above 100 MeV is  $\sim  6\times10^{-7} \rm  cm^{-2} s ^{-1 }sr^{-1}$ \cite{2012JCAP...02..012F} ,which we safely extrapolate to a flux above 50 MeV of $3\times 10^{-6} \rm  cm^{-2} s ^{-1 }sr^{-1}$
or $4\times10^{-5} \rm cm^{-2 } s^ {-1}$, against  $\sim 1  \rm cm^{-2 }s^{-1}$  deduced from the calculated (still unobserved) antineutrino background  \cite{2017JCAP...11..031P}.
So $f_{KK}$ is about $4 \times10^{-5}$. The exact value is not of crucial importance since M* is proportional to $f_{KK}^{1/5} $  
or $f_{KK}^{1/6}  $ for n = 3 and 4, thus order of magnitude estimates  are sufficient. 

Consequently, $f_{KK} $ is $\sim$250 times lower than that assumed in \cite{2003PhRvD..67l5008H,2004PhRvD..69b9901H}
($10^{-2}$ ). Using values from Table VI in the same reference, we derive the limits M*$ >$ 8.0 TeV and 1.1 TeV for n = 3 and 4. These limits can be compared to those obtained at the LHC by the ATLAS and CMS collaborations searching for direct production of gravitons together with one or several jets, namely ATLAS: 2.1 TeV and 1.6 TeV  \cite{Aaboud}
and CMS: 2.5 and 1.9 TeV \cite{Sirunyan}
for n=3 and n=4 respectively, or by CMS searching for a contribution to dijet production through virtual graviton exchange: 4.0 TeV and 3.0 TeV
\cite{EPJC...78..891S}
 for n=3 and n=4 respectively. 
The constraints from this analysis are found to be of the same order of magnitude as those from the LHC, more stringent in the case n=3 and less so  in the case n=4 .
Indeed the limits derived from the GRB are very robust. A factor of 10 uncertainty on $f_{KK}$ translates only into a factor 1.4 uncertainty on M* (min) for n = 4. Thus this method liberates us from the tyranny of precision. 

\subsection{Approximations}
 Concerning the  energy range, the gamma ray emission of gravitons is limited to about 300 MeV 
 whereas we have considered the whole photon spectrum for simplicity. This makes little difference due to the steepness of the GRB (slope of $-2.3$). 

 With regard to undecayed gravitons, the lifetime of gravitons is very long, about $(100\  \rm MeV/m)^3$  Gyr, with m being their mean mass (68 and 95 MeV for n = 3 and 4 (\cite{2003PhRvD..67l5008H},
 Table 2). Their lifetime is of the order of 1 Gyr for n = 3 and 4, corresponding to a redshift about 0.1. From the calculated extragalactic antineutrino background which peaks around 5 MeV \cite{2018JCAP...05..066M,2017JCAP...11..031P}  whereas the mean emission energy of SN antineutrinos is about 15 MeV (weighted average over the Initial Mass Function \cite{2018MNRAS.475.1363H},
 we  deduce that the maximum contribution of SN to the background is at $z \sim 2,$ corresponding to the maximum in the cosmic star formation rate. 
 
 \subsection{Uncertainties} 
 There are many sources of uncertainties, 
 the principal one being that of the calculated neutrino flux, in turn due to that of the present core collapse SN rate \cite{2018JCAP...05..066M,2017JCAP...11..031P},
 by a factor  of the order of 2. 
 
 The uncertainty in the mean temperature during core collapse is much less dramatic than in the case of individual neutron stars \cite{2003PhRvD..67l5008H}
 used to constrain M*. Here T = 30 MeV is the minimum value of the mean temperature during core collapse. The bounds are quite insensitive to the assumed temperature of the emitting medium because, for a larger temperature, the average energy of the emitted KK states increases, leading to a decrease in their total number, and at the same time, the energy of the decay photons is distributed over a broader range of energies, further decreasing the differential flux.
  Therefore, the predicted photon flux is lower, but extends to larger energies. On the other hand, the measured gamma ray flux falls approximately as $E_\gamma^{-2}$ canceling the previous effect. Finally, note that the bounds derived in \cite{2004PhRvL..92k1102C}, 
 based on the gamma ray emission of the (old) galactic bulge, especially for n = 3 should be reassessed since most of the gravitons would have decayed, lessening the flux from the bulge.
 On the other hand, one should also consider not the whole gamma ray emission of the bulge but the residual emission left by photons from graviton decays, which is difficult to ascertain. Thus the present limit on n = 3 replaces the previous one. Note once again that constraints on ED from the observations of gravitational waves \cite{2018JCAP...07..048P},
 indicating that gravitational  and electromagnetic waves propagate at the same velocity, do not apply in our case since ED are compact (their radius is much less than the wavelengths of gravitational waves).


\section{GALACTIC GAMMA RAYS AND EXTRA DIMENSIONS}
Assuming the ADD scenario, we calculate the gamma ray luminosity of the Milky Way in erg s$^{-1}$, induced by the decays of gravitons produced by core collapse and trapped around compact objects, neutrons stars and black holes (remnants of successful and failed SN).
We then compare this to the space left by gamma rays produced by cosmic rays and young pulsars  i.e. the residuals (calculated and observed) of the galactic emission above 100 MeV.
\footnote {Millisecond pulsars are too old to contribute, all gravitons should have decayed.}

The influential parameters are:
\begin{itemize}
\item $r_{SN}$, the SN birth rate and $r_{BH}$ the  black hole(BH) birth rate, taken as  2/century and 0.15/century, respectively
\item  $ M_{BH}$, the mean mass of stellar BH  $\sim 10\rm M_\odot$
\item $E_{grav }$, the total (gravitational) energy released in core collapse, ($\sim M^2/R$, $3\times10^{53}$ and $4.5 \times 10^{54}$ erg, respectively;
SN and BH contribute almost equally, since the lower number is compensated by a higher emissivity. 
\item $f_{kk }$ the fraction of energy released in the form of gravitons, which is the parameter of interest since $M^\ast \sim f_{KK}^{-1/(n+2)}$
\item$f_{trap}$, the fraction of gravitons trapped by the compact remnant,  the same being  true for NS and BH, taken as $ 1/3$ for n =3, 4, 5 (\cite{2003PhRvD..67l5008H},
table II)
\item$\tau_\gamma$, the gamma ray lifetime and  $\tau$, the mean total lifetime, which is approximated as  $ \tau_\gamma/4$  taking into account other decay channels  (electron/positron and neutrino/antineutrino pairs, see eq. 44 in \cite{2003PhRvD..67l5008H}, with $\tau_\gamma =\tau_e/2=\tau_\nu$.
For $n > 3,$  $\tau$ is relatively short ($<$100 Myr). Over this time-scale, the production rate of gravitons, proportional to the SN rate, can be considered as approximately  constant. Thus a steady state situation prevails. 

\end{itemize}

Each second, an energy of  $1.3 \times 10^{44}$  erg times $f_{KK}$  is injected into the galaxy in the form of gravitons by SN explosions. Most of them fly away and are lost from the galaxy since only a fraction $  f_{trap} \approx 1/3$ remain trapped around NS and BH.  
 The steady state energy in the form of gravitons confined permanently around neutron stars is $q\tau$ with 
$q= f_{KK}f_{trap}\times 6\times10^{53}$  ergs times the SN rate of 2 per century, or $4\times 10^{44 } $ erg s$^{-1}$.

For  $f_{KK}= 1,$   the gamma ray luminosity is $L_\gamma =2  f_{trap}  q\tau/\tau_\gamma,$ since one graviton produces $\sim  q/6$, almost independently of the number n of ED 
 i.e. $6.7\times 10^{43} \rm  erg s^{-1}.$
This figure is to be compared to the residuals left by other processes (cosmic rays and possibly dark matter), say $5\times10^{37} \rm  erg s^{-1}$, i.e. 10\% of the emission observed by Fermi-LAT \cite{2012JCAP...02..012F}.
Our final result is now: $f_{KK}  < 1.3 \times 10^{-6}$. This compares to the careful analysis of the Fermi-LAT group of the lack of gamma emission of selected individual  neutron  stars, which leads to $f_{KK}=8.7\times  10^{-3}$ for n=3 and $7.4\times  10^{-4}$ for n=4.

\section{Conclusion}
In the framework of the ADD extra-dimensional model for gravity at the TeV scale, we have revisited the constraints on the maximum fraction of energy emitted in the form of massive (Kaluza-Klein) gravitons by core collapse supernovae, coming from the extragalactic gamma-ray background and from  residual galactic emission. From these constraints, we derive limits on the mass parameter M* of the model for a number of extra dimensions of 3 and 4. 
Table \ref{tab:summary} summarizes the limits on  M* obtained in this work and in previous references mentioned in the text.
\begin{table}
\begin{tabular}{lcccccc}
\hline\hline

  &This work&This work&HR 2003&Fermi& LHC \\
  & diffuse  & galactic  &HR 2004& -LAT &  CMS \\
  &gamma &gamma&  & 2012 & virtual \\
  &  rays & rays &  &  &graviton\\\hline
n=3 & 8.0 & 16 & 2.65  & 5.3 & 4.0\\
n=4 &  1.1  & 1.9 & 0.43  & 0.74& 3.0\\
\hline\hline
\end{tabular}
\caption{Limits on M* in TeV, for this work and previous references in the text.}
\label{tab:summary}
\end{table}

\section*{Acknowledgements}

We thank Elisabeth Vangioni for useful discussions. 

\bibliographystyle{mn2e}


\end{document}